\documentstyle[ltwol,epsfig]{article}

\def\nuebar{\bar{\nu_e}}
\def\dm2{\rm{\Delta m^2}}

\def\s2tw{\rm{ sin ^2 \theta _W }}
\def\am241{\rm{ ^{241} Am }}
\def\u238{\rm{ ^{238} U }}
\def\th232{\rm{ ^{232} Th }}
\def\k40{\rm{ ^{40} K }}

\begin{document}


\title{
\hfill   {\bf AS-TEXONO/01-02}\\
  \hfill      \today \\[2ex]
NEUTRINO EXPERIMENTS : HIGHLIGHTS
}

\author{HENRY TSZ-KING WONG}

\address{
Institute of Physics, Academia Sinica, Nankang 11529, Taipei, Taiwan.\\
E-mail: htwong@phys.sinica.edu.tw
}


\twocolumn[\maketitle\abstracts{ 
This article consists of two parts.
The first section presents  
the highlights on the 
goals of neutrino physics,
status of the current neutrino
experiments and future directions
and program.
The second section describes the theme,
program and research efforts for the 
TEXONO
Collaboration among scientists
from Taiwan and China.
}]

\section{Introduction}

With the strong evidence 
presented by the Super-Kamiokande experiment on
neutrino oscillations~\cite{superk}, there  are intense
world-wide efforts to pursue the next-generation
of neutrino experiments. 

The aims of the first section of this
article is to ``set the stage''
for students and researchers not in the field
by summarizing the key ingredients and highlights
of the goals, status and future directions in
neutrino physics experiments.
It is not meant to be a comprehensive lecture or
detailed review article.
Interested readers can pursue the details from
the listed references on textbook accounts~\cite{boehm},
latest status~\cite{conf} and Web-links~\cite{weblink}.

The second part of this article presents an
account of the research program of the 
TEXONO Collaboration.

\section{Neutrino Physics Experiments}

\subsection{Why Neutrino Physics}

Neutrino exists $-$ and exists in large quantities in
the Universe, comparable in number density to the photon.
It is known that there are three flavors
of light neutrino coupled
via weak interaction to the Z gauge boson.
Yet the fundamental properties :
(1) masses, denoted by $\rm{m_i}$ for mass eigenstate $\rm{\nu_i}$, and
(2) mixings, denoted by $\rm{U_{\it{l}i}}$ for mixing matrix elements
between flavor eigenstate $\nu_l$ and mass eigenstate
$\rm{\nu_i}$  
(alternatively by $\rm{\theta_{ij}}$ for mixing angles
between mass eigenstates $\rm{\nu_i}$ and $\rm{\nu_j}$),
remains unknown or at least 
not accurately known enough.

In field theory language, this translates to
the crucial question on 
the structure (or even the possible existence) of a
``neutrino mass term'' L($\nu$-mass) in the total
Lagrangian. Standard Model sets this to be identically
zero, but without any compelling reasons $-$ in
contrast to the massless-ness of the photons being dictated
by gauge invariance.
The detailed structures and values
of this term can reveal much about
the Grand Unified Theories.

At the large length-scale frontier,  
neutrino mass is related to
the composition and structural evolution of
the Universe. 
Neutrino has been a candidate
of Dark Matter~\cite{conf}: in fact it is the only
candidate within the list that is proven to exist.

\begin{figure}
\center
\epsfig{figure=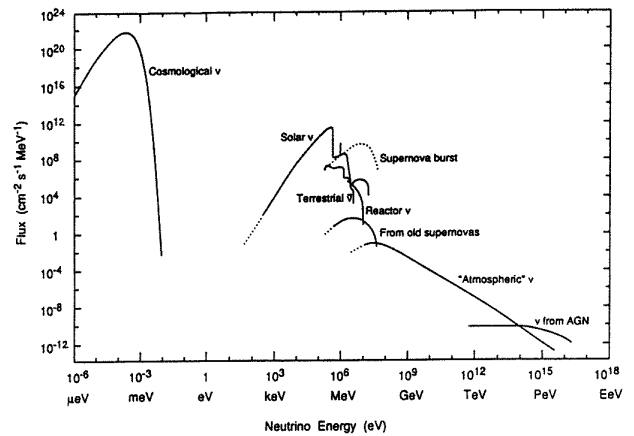,height=2.7in}
\caption{
The expected neutrino spectra from
various celestial and terrestrial sources.
Neutrinos from man-made accelerators, typically
at the range of 10~MeV to 100~GeV, are not
shown here since accelerator parameters differ.
}
\label{nusource}
\end{figure}

Experimentally, the probing of L($\nu$-mass)
is carried out by studying various processes
related to neutrino masses and mixings, such as direct mass
measurement through the distortion of $\beta$-spectra,
neutrinoless double beta decays, neutrino oscillations,
neutrino magnetic moments, neutrino decays ..... and
so on. These investigations are realized by a wide
spectrum of experimental techniques spanned over several
decades of energy scale with different neutrino sources. 
The expected neutrino spectrum due to terrestrial
and astrophysical sources are shown in Figure~\ref{nusource}.

In addition, neutrino has been used as probe (as ``beam''
from accelerators and reactors and even astrophysical
sources) to study electroweak physics,
QCD, structure function physics,  nuclear physics,
and to provide otherwise inaccessible information on
the interior of stars and supernovae.

Therefore, the study of 
neutrino physics and the implications
of the results connect many disciplines together,
from particle physics to nuclear physics to astrophysics
to  cosmology.

\subsection{Current Status and Interpretations}

Neutrino interactions are characterized by cross-sections
at the weak scale (100~fb at 100~GeV to $< 10^{-4}$~fb      
at 1~MeV). As an illustration, the mean free path in water
for $\nuebar$ from power reactors at the typical
energy of 2~MeV  is 250 light years!
The central challenge to neutrino experiments is
therefore on how to beat this small cross section
and/or slow decay rate. Usually massive detectors
are necessary to compensate by their
large target size. 
Then the issue becomes how to keep the cost and 
background low.

\begin{figure}
\center
\epsfig{figure=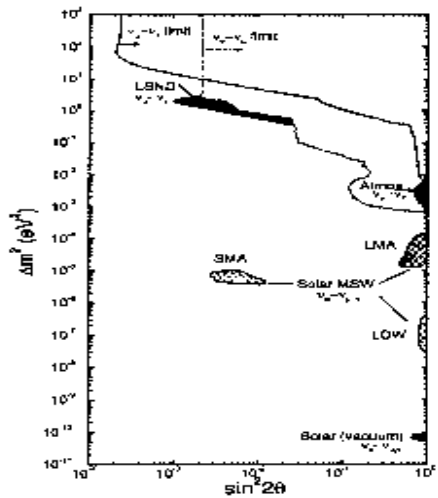,width=6cm,height=2.7in}
\caption{
A summary of results of neutrino oscillation
experiments in ($\rm{\Delta m^2}$,$\rm{ sin ^2 2 \theta }$)
parameter space.
The allowed regions are shaded.
}
\label{nuosc}
\end{figure}

After half a century of ingenious experiments
since the experimental discovery of the neutrinos
by Cowan and Reines,
there are several results which may indicate
the existence of neutrino masses, and hence
physics beyond the Standard Model. All these
results are based on experimental searches
of neutrino oscillation, a 
quantum-mechanical effect which 
allows neutrino to transform from one flavor
eigenstate to another as it traverses in space.
This process depends on 
the mass difference 
($\rm{\Delta m^2 = | m_i^2 - m_j^2 | }$)
rather than the absolute mass,
resulting in enhanced
sensitivities.  
A simplified summary of the results
of neutrino oscillation experiments
is shown in Figure~\ref{nuosc}.

The ``allowed regions'' are due to anomalous
results from experiments in:
\begin{enumerate}
\item 
{\bf Atmospheric Neutrinos:}\\
Data from 
the Super-Kamiokande experiments~\cite{superk},
supported by other experiments,
indicates a smaller  
$\rm{ ( \nu_{\mu} + \bar{\nu_{\mu}} ) /
( \nu_{e} + \bar{\nu_{e}} ) }$ 
ratio
than would be expected from propagation
models of cosmic-ray showering. 
The ``smoking gun'' for new physics is 
that the 
deficit has a statistically strong
dependence with the zenith angle, meaning
the effect depends on
the propagation distance of the neutrinos 
from the production point to the detector.
The combined fit supports a scenario
of $\rm{\nu_{\mu} \rightarrow \nu_{\tau}}$
oscillation. 
\item
{\bf Solar Neutrinos:}\\
All solar neutrinos to date reported 
an observed solar neutrino flux less
than the predictions of Standard Solar
Model. The deficit is different
among the experiments, suggesting an
energy dependence which is difficult
to be explained by standard solar models.
However, within an individual experiment, the potential
``smoking gun'' effects (day-night variation,
seasonal variation, spectral distortion) 
are absent or statistically weak.
The combined data favors 
neutrino oscillations of $\rm{\nu_{e}}$ to
another species, active or sterile:
$\rm{\nu_{e} \rightarrow \nu_{x}}$,
either in vacuum or due to matter-enhanced
oscillation (the ``MSW'' Effect) in the Sun.
\item
{\bf LSND Anomaly:}\\
The LSND experiment with accelerator neutrinos
reported unexpected excess of $\rm{\nu_{e}}$ 
and $\rm{\bar{\nu_{e}}}$ in a 
$\rm{  \nu_{\mu} + \bar{\nu_{\mu}} }$ beam,
which can be explained by 
$\rm{\nu_{\mu} \rightarrow \nu_{e}}$
oscillation. The results are yet to
be reproduced (or totally excluded)
by other experiments.
\end{enumerate}

If all experimental results are correct and
to be explained by neutrino oscillations, one
must incorporate a fourth-generation sterile
neutrino. If one takes the conservative approach
that LSND results must be reproduced by an
independent experiment
before they are incorporated into
the theoretical framework, then the 
favored scenario would be  a three-family
scheme with:
\begin{itemize}
\item
{\bf Neutrino Mass Differences:}\\
Atmospheric neutrino oscillation is driven by
$\rm{\Delta m^2 _{23}  \sim   10^{-2}  - 10^{-3} ~ eV^2 }$
which is much bigger than the scale which
drives solar neutrino oscillation at
$\rm{ \Delta m^2 _{12}  \sim  10^{-4}  - 10^{-5} ~  eV^2 }$
due to matter-enhanced oscillation,
as depicted in Figure~\ref{mhier}. The
sign of $\rm{\Delta m^2 _{23}}$   remains undetermined.
\begin{figure}
\centerline{
\epsfig{file=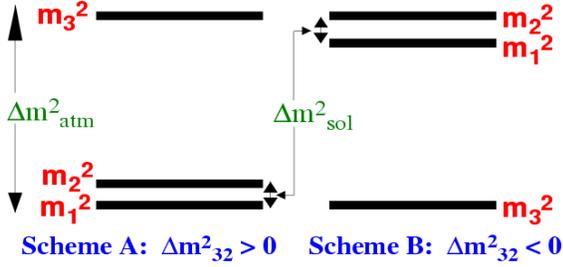,width=8cm,height=4cm}
}
\caption{
The favored hierarchical structures of the neutrino
mass matrix to explain atmospheric and solar neutrino
results.
}
\label{mhier}
\end{figure}
\item
{\bf Mixing Angles:}\\
Surprisingly when compared to the quark sector,
the atmospheric
neutrino data supports strongly large mixing angles
(order of 1) for $\rm{sin ^2 2 \theta _{23}}$.
Combining all data from solar neutrinos also 
statistically favors large $\rm{sin ^2 2 \theta _{12}}$.
The 1-km baseline reactor neutrino experiments
set limits on $\rm{ sin ^2 2 \theta _{13} } <  0.1 $.
\end{itemize}

\subsection{Future Experimental Program}

Among the various neutrino sources
depicted in Figure~\ref{nusource}, only
a relatively small window from $\sim$1~MeV to
$\sim$100~GeV is detectable by present techniques.
The future of neutrino experiments
will therefore evolve along various  directions:

\subsubsection{(I) Further Exploration of the Measured Window}
\begin{itemize}
\item {\bf Long Baseline (LBL) Neutrino Oscillation Experiments:}\\
There is a running experiment with
accelerator neutrinos
produced by ``proton-on-target'' (PoT)
from KEK to the Super-Kamiokande detector (250~km), while two
others (both 730~km)
in construction: Fermilab to MINOS experiment at Soudan,
and CERN to ICANOE and OPERA experiments at Gran Sasso.
The goals will be to reproduce and improve on the 
$\rm{\nu_{atm}}$ 
parameters, to observe
the E/L oscillation pattern, and to detect
$\nu_{\tau}$ appearance explicit by observing
$\nu_{\tau}$ charged-current interactions leading
to the production of $\tau$-lepton.\\
In addition, there are two experiments, KamLAND
and BOREXINO, with capabilities of performing
LBL experiments with reactor neutrinos from
power plants $\sim$100~km away. Their goals will 
be the probe the ``Large Mixing Angle'' MSW solution
to the solar neutrino data.
\item {\bf Detection of Weak/Rare Signals:}\\ 
Various big underground experiments are
sensitive to neutrinos from supernovae, with
the hope of detecting thousands of events from
the next supernova, as compared to the 20~events
from SN1987a. The LBL reactor neutrino experiments
will also try to observe ``terrestrial neutrinos''
produced by the radioactive (mainly $^{232}$Th
and $^{238}$U series) from the Earth's crust.
\item {\bf Further Double Beta Decay Searches:}\\
Neutrinoless Double Beta Decay is sensitive
to the absolute scale of $\rm{m_i^2}$ rather
then $\rm{\Delta m^2 _{ij}}$. Many R\&D projects
are underway to try to achieve the interesting
range of $\rm{m_3 \sim 0.1 - 0.01}$ which
is a scenario suggested by atmospheric neutrino
results.
\item {\bf Neutrino Factories from Muon Storage
Rings:}\\
There are intense pilot efforts to study the
feasibilities of performing LBL experiments
with a ``Neutrino Factory'' where
the neutrinos are produced by muons decay
in a muon storage ring.
Unlike conventional PoT neutrino beam,
$\mu$-decay gives beams which are 
selectable in $\rm{\nu_e}$, \rm${\bar{\nu_e}}$
$\rm{\nu_{\mu}}$ and $\rm{\bar{\nu_{\mu}}}$,
and with well-known spectra and compositions.
Coupled to the high luminosity and small beam
size, they can be powerful tools to study
neutrino physics. The goals for the LBL program,
where the baseline will have to be the
1000-10000~km range,
would be to do precision measurements on
the $\rm{\nu_{atm}}$ parameters, to 
probe $\rm{sin ^2 2 \theta_{13}}$ to O($10^{-4}$) 
(from $\rm{\nu_{e} \rightarrow \nu_{\mu}}$),
to determine the signs of $\rm{\Delta m^2 _{23}}$
(from the asymmetric $\rm{\nu_e / \bar{\nu_e}}$ matter effects),
and to probe CP-violation in the lepton sector
(comparing $\rm{\nu_{e} \rightarrow \nu_{\mu}}$ \&
$\rm{\bar{\nu_{e}} \rightarrow \bar{\nu_{\mu}}}$).
There are many technical challenges 
which must be addressed towards the
realization of such projects.
\end{itemize}

\subsubsection{(II) Extension of Detection Capabilities:}

\begin{itemize}
\item
{\bf High Energy Frontiers:}\\
There are several ``neutrino telescope'' projects
(Lake Baikal, AMANDA, NESTOR, ANTARES) whose
objective is towards the construction
of an eventual ``km$^3$'' detector. 
The scientific goals are (a) to identify and
understand the astrophysics of high-energy (TeV to PeV)
neutrino sources from active galactic nuclei, gamma-ray
bursts, neutron stars and other astrophysical
objects, and (b) to use these high-energy neutrinos
for neutrino physics like very long baseline
studies.
\item
{\bf Low Energy Frontiers:}\\
There are a host of new solar neutrino
experiments: SNO is taking data while
Borexino and KamLAND are under construction,
as well as many R\&D ideas and projects.
Their goals are to measure the
solar neutrino spectrum and particularly
those from the dominant sub-MeV pp and 
$^7$Be neutrinos, and to study the 
details of spectral shape, day-night 
and seasonal variations, as well as the 
neutral-current to charged-current ratios, 
so as to identify a unique solutioin to 
``solar neutrino problem'', 
and to measure 
the ($\rm{\Delta m^2, sin ^2 2 \theta}$) parameters. 
The low-energy solar neutrino spectrum
also provides a probe to the other inaccessible
physics at the interior of the Sun.
\\ Weakly Interacting Massive Particles (WIMP) 
are candidates for Dark Matter. 
Their experimental searches also employ the 
techniques of low-background low-energy
experiments. Many experiments are running or being
planned, based on crystal scintillators,
cryogenic detectors and other techniques.\\
Finally, the relic ``Big Bang'' neutrino, the
counterpart to the 2.7~K cosmic microwave photon background (CMB),
has large number density (110~$\rm{cm^{-3}}$ for Majorana
neutrinos, comparing 
to 411~$\rm{cm^{-3}}$ 
for CMB) but extremely small cross sections due to
the meV energy scale at an effective temperature of 1.9~K.
The relic neutrinos decouples from matter
at a much earlier time (1~s) than the CMB (3$\times 10^5$~years),
and hence are, in principle, better probes to the
early Universe.
A demonstration of its existence and a measurement
of its density is a subject of extraordinary
importance. Though there is no
realistic proposals on how to detect them,
it follows the traditions of 
offering a highly rewarding challenge to
and pushing the ingenuity of 
neutrino experimentalists.
\end{itemize}

\section{Research Program of the TEXONO Collaboration}

Since 1997, the 
TEXONO\footnote{{\bf T}aiwan {\bf EX}periment {\bf O}n {\bf N}eutrin{\bf O}}
Collaboration has been built up to
initiate and pursue an experimental
program in Neutrino and Astroparticle Physics~\cite{start}.
By the end of  2000, the Collaboration comprises
more than 40 research scientists from
major institutes/universities
in Taiwan (Academia Sinica$^{\dagger}$, Chung-Kuo
Institute of Technology, Institute of Nuclear
Energy Research, National Taiwan University, National Tsing Hua
University, and Kuo-Sheng Nuclear Power Station),
Mainland China (Institute of High Energy Physics$^{\dagger}$,
Institute of Atomic Energy$^{\dagger}$,
Institute of Radiation Protection,
Nanjing University)
and the United States (University of Maryland),
with AS, IHEP and IAE (with $^{\dagger}$)
being the leading groups.
It is the first research collaboration 
of this size and magnitude.
among Taiwanese and Mainland Chinese scientists 

The research program~\cite{program} is based on the
the unexplored and unexploited theme of adopting
the scintillating crystal detector techniques
for low-energy low-background experiments
in Neutrino and Astroparticle Physics~\cite{prospects}.
The ``Flagship'' experiment~\cite{expt} is based
on CsI(Tl) crystals placed near  the core of 
the Kuo-Sheng Nuclear Power Station (KSNPS)
at the northern shore of Taiwan to
study low energy neutrino interactions.
It is the first particle physics experiment
performed in Taiwan where local scientists
are taking up  major roles and responsibilities
in all aspects of its operation:
conception, formulation, design, prototype
studies, construction, commissioning,
as well as data taking and analysis.

In parallel to the flagship reactor experiment,
various R\&D efforts coherent with the
theme are initiated and pursued.

\subsection{Scintillating Crystal Detector for
Low-Energy Experiments}
\label{sect::det}

One of the major directions and experimental
challenges in neutrino physics~\cite{conf}
is to extend the measurement capabilities to
the sub-MeV range for the detection of
the p-p and $^7$Be solar neutrinos, Dark Matter
searches and other topics.
For instance, while
high energy (GeV) neutrino beams from accelerators
have been very productive in the
investigations of electroweak,
QCD and structure function
physics,
the use of low energy (MeV) neutrino
as a probe to study particle and nuclear physics
has not been well explored.
Nuclear power reactors  are abundant source of
electron anti-neutrinos ($\nuebar$) at the MeV range
and therefore provide a convenient laboratory
for these studies.

On the detector technology fronts,
large water Cerenkov and liquid scintillator detectors
have been successfully used in
neutrino and astro-particle physics experiments.
New detector technology must be explored to
open new windows of opportunities.
Crystal scintillators may be well-suited
to be adopted for low background experiments
at the keV-MeV range~\cite{prospects}.
They have been
widely used as electromagnetic calorimeters in
high energy physics~\cite{emcalo}, 
as well as in medical and
security imaging
and in the oil-extraction industry.
There are matured experience
in constructing and operating scintillating
crystal detectors to the mass range of 100~tons.

\begin{figure}
\centerline{
\epsfig{file=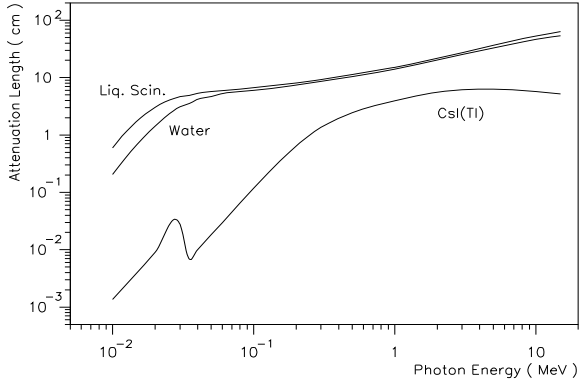,height=2in,width=8cm}
}
\caption{
The attenuation length, as defined by the interactions
that lead to a loss of energy in the media,
for photons at different energies,
for CsI(Tl), water, and liquid scintillator.
}
\label{attlen}
\end{figure}

This technique offer many potential merits
for low-energy low-background experiments.
In particular, they are usually made of 
high-Z materials which provide
strong attenuation for $\gamma$'s of energy less
than 500~keV, as indicated in the photon
attenuation plot in Figure~\ref{attlen}.
As an illustration,
10~cm of CsI(Tl)
has the same attenuating power as 5.6~m of liquid
scintillators at $\gamma$-energy of 100~keV.
Consequently, it is possible to realize
a compact detector design with
minimal passive materials equipped
with efficient active veto and passive shielding.
Externally originated photons in this energy range
from ambient radioactivity or from surrounding
equipment cannot penetrate into the fiducial volume.
Therefore, the dominating contribution to the
experimental sensitivities is expected to be
from the internal background in the crystal itself,
either due to intrinsic radioactivity or cosmic-induced
long-lived isotopes, both of which can be identified
and measured such that the associated
background can be subtracted off accordingly.
The experimental challenges are focussed on
the understanding and control of the {\it internal background}.
Pioneering efforts have already been  made
with NaI(Tl) crystals for Dark Matter searches~\cite{nai}.

\subsection{Flagship Experiment with Reactor Neutrinos}

\begin{figure}
\center
\epsfig{figure=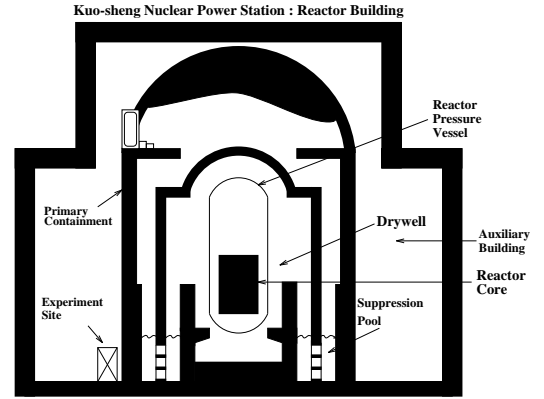,height=2.7in,angle=270}
\caption{
Schematic side view, not drawn to scale,
of the Kuo-sheng Nuclear Power Station
Reactor Building,
indicating the experimental site.
The reactor core-detector distance is about
28~m.
}
\label{ksnpssite}
\end{figure}

An experiment towards a 500~kg 
CsI(Tl) scintillating crystal detector to
be placed at a distance of 28~m from
a core in KSNPS is
under construction
to  study various
neutrino interactions at the keV-MeV range~\cite{expt},
and to establish and explore the general techniques
for other low-energy low-background applications.
The layout of the experimental site is shown
in Figure~\ref{ksnpssite}

The physics objectives of the experiment  are
to improve on or to explore the subjects of:
(1) Neutrino-electron scatterings, which
is a fundamental interaction 
providing information on the electro-weak parameters
($\rm{ g_V , ~ g_A , ~ and ~ sin ^2 \theta_W }$),
and are sensitive to
small neutrino magnetic moments ($\mu{\nu}$)
and the mean square charge radius,
as well as the W-Z interference effects;
(2) Neutrino interactions on nuclei such as
$^{133}$Cs and $^{127}$I, and
in particular the neutral current excitation (NCEX) channels where
the signatures are gamma-lines at the characteristic energies $-$
their cross-sections are related to the physics of
nucleon structure, and are essential to explore the
scenario of applying these interaction channels 
in the detection of low energy solar and supernova
neutrinos;
(3) Matter effects on neutrinos, since this
is the first big high-Z detector built for reactor
neutrino studies.

\begin{figure}
\center
\epsfig{figure=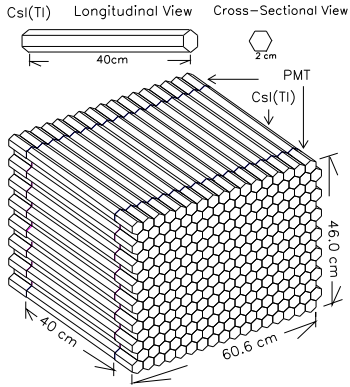,height=2.0in}
\caption{
Schematic drawings of the CsI(Tl)
target configuration. 
}
\label{csitarget}
\end{figure}

To fully exploit the advantageous features
of the scintillating crystal approach
in low-energy low-background experiments,
the experimental configuration
should enable the definition of a fiducial
volume with a surrounding active 4$\pi$-veto,
and minimal passive materials. 
This is realized by a design shown
schematically in Figure~\ref{csitarget}.
One CsI(Tl) crystal unit
consists of a hexagonal-shaped cross-section with 2~cm
side and a length 20~cm, giving a mass of 0.94~kg.
Two such units are glued optically
at one end to form a module.
The modules will be 
installed in stages towards an eventual
$17 \times 15$ matrix configuration.
The light output
are read out
at both ends  by
custom-designed 29~mm diameter
photo-multipliers (PMTs) with low-activity glass, whose
signals will pass through amplifiers and shapers 
to be digitized by 20~MHz FADCs~\cite{electronics}.
The sum and difference of the PMT signals gives information
on the energy and the longitudinal position of
the events, respectively.
The passive shieldings consist of, from inside out,
5~cm of copper, 25~cm of boron-loaded polyethylene, 5~cm
of steel, 15~cm of lead and finally plastic scintillators
as cosmic-ray veto. The target is housed in a nitrogen
environment to prevent background events due to the
diffusion of the radioactive radon gas.

\begin{figure}
\center
\epsfig{file=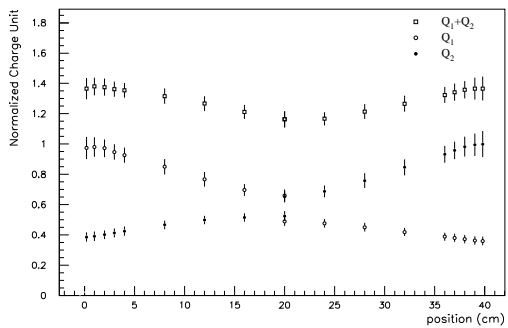,height=2.0in}
\caption{
The measured variations of
Q$_1$, Q$_2$ and $\rm{Q_{tot}= Q_1 + Q_2}$ along
the longitudinal position of the crystal module.
The charge unit is normalized to unity for both
Q$_1$ and Q$_2$ at their respective ends.
}
\label{qvsz}
\end{figure}

Extensive measurements on the crystal prototype
modules have been performed~\cite{proto}.
The response is depicted
in Figure~\ref{qvsz}, showing the variation
of collected light for $\rm{Q_1}$, $\rm{Q_2}$
and $\rm{Q_{tot}}$ as a function of position
within one crystal module.
The error bars denote the FWHM width of the
$^{137}$Cs photo-peaks. The discontinuity at L=20~cm
is due to the optical mis-match between the
glue (n=1.5) and the CsI(Tl) crystal (n=1.8).
It can be seen that
$\rm{Q_{tot}}$ is only weakly 
dependent of the position
and a 10\% FWHM energy 
resolution is achieved at 660~keV.
The detection threshold (where signals
are measured at both PMTs) is $<$20~keV.
The longitudinal position can be obtained
by considering the variation of the ratio
$\rm { R = (  Q_1 - Q_2 ) / (  Q_1 + Q_2 ) }$
along the crystal.
Resolutions of 2~cm and 3.5~cm at
660~keV and 200~keV, respectively, have been demonstrated.

In addition, CsI(Tl) provides powerful
pulse shape discrimination capabilities
to differentiate $\gamma$/e from $\alpha$ events,
with an excellent separation of $>$99\% above 500~keV.
The light output for $\alpha$'s in CsI(Tl) is quenched
less than that in liquid scintillators.
The absence of multiple $\alpha$-peaks 
above 3~MeV in the prototype measurements
suggests that a $^{238}$U and $^{232}$Th
concentration 
(assuming equilibrium)
of $< 10^{-12}$~g/g can be achieved.

By early 2001, 
after three and a half years of preparatory
efforts, the
``Kuo-Sheng Neutrino Laboratory'',
equipped with
flexibly-designed shieldings,
cosmic-ray 4-$\pi$ active veto, complete electronics which
allow full digitization of multi-channel detectors
for a 10~ms long duration,
data acquisition and monitoring systems, as well
as remote assess capabilities,
was formulated, designed, constructed and commissioned.
Prototype CsI(Tl) detector has been operating
in the home-base laboratory.
The first Reactor ON/OFF data taking 
period will be based on a 1~kg low-background germanium
detector with 100~kg of CsI(Tl) running
in conjunction.

\subsection{R\&D Projects}

Various projects with
stronger R\&D flavors are
proceeding in parallel to the flagship
reactor experiment.
A feasibility study of using boron-loaded
liquid scintillator for the detection of
$\nuebar$ has completed~\cite{liqscin}.
The cases of using of GSO~\cite{gso} and 
LiI(Eu)~\cite{lii} as well as
Yb-based scintillating crystals
for low energy solar neutrinos are explored.
The adaptations of CsI(Tl) crystal scintillator
for Dark Matter 
WIMP searches are studied, which includes
a neutron beam measurement
to study the response due to nuclear recoils ~\cite{ncalib}.

A generic and convenient technique to measure 
trace concentration of radio-isotopes  
in sample materials is of great importance
to the advance of 
low-energy low-background experiments.
A R\&D program is initiated
to adapt the techniques
of Accelerator Mass Spectrometry(AMS)
with the  established facilities
at the 13~MV TANDEM accelerator
at IAE~\cite{ciaeams}.
The goals are to device methods
to measure $^{238}$U,
$^{232}$Th, $^{87}$Rb, $^{40}$K 
in liquid and crystal scintillators
beyond the present capabilities~\cite{amsradio}.

Complementary to these physics-oriented program
are detector R\&D  efforts.
Techniques to grow CsI(Tl) mono-crystal
of length 40~cm, the longest in the world,
have been developed and are deployed in
the production for future batches.

\section{Outlook}

Neutrino physics and astrophysics will remain
a central subject in experimental particle physics
in the coming decade and beyond. There
are room for ground-breaking technical innovations -
as well as potentials for surprises in the scientific
results.

A Taiwan, Mainland China and U.S.A.
collaboration has been built up 
with the goal of playing a major
role in this field. 
It is 
the first generation collaborative efforts
in large-scale basic research between scientists
from Taiwan and Mainland China.
The technical strength and scientific
connections of the Collaboration
are expanding and consolidating.
The flagship experiment is to
perform the first-ever particle
physics experiment in Taiwan.
Many R\&D projects are being pursued.

The importance of the implications and 
outcomes of the experiment and 
experience will
lie besides, if not beyond, neutrino physics.

\section*{Acknowledgments}

The author is grateful to the members,
technical staff and industrial partners
of TEXONO Collaboration, as well as 
the concerned colleagues
for the many contributions which ``make 
things happen'' in such a short period
of time.
Funding are
provided by the National Science Council, Taiwan
and 
the National Science Foundation, China, as well
as from the operational funds of the 
collaborating institutes.

\end{document}